\documentclass{fic-l}
\usepackage{graphicx}
\usepackage{amscd}
\usepackage{amsmath}
\usepackage{amsfonts}
\usepackage{amssymb}

\theoremstyle{plain}

\numberwithin{equation}{section}

\begin{document}
\title{Space- and Time-Like Superselection Rules in Conformal Quantum Field Theory}
\author{Bert Schroer}
\address{Prof. emeritus, Institut f\"{u}r Theoretische Physik, FU-Berlin, presently
CBPF Rio de Janeiro}
\email{schroer@cbpf.br}
\date{September 2000, \  {\small to appear in the proceedings of the Siena
conference on ''MATHEMATICAL PHYSICS IN MATHEMATICS AND IN PHYSICS'', SIENA,
19 - 25 JUNE 2000}}
\date{to be published by the Fields Institute}
\date{September 2000, \  {\small to appear in the proceedings of the Siena
conference on ''MATHEMATICAL PHYSICS IN MATHEMATICS AND IN PHYSICS'', SIENA,
19 - 25 JUNE 2000}}
\date{to be published by the Fields Institute}
\dedicatory{Dedicated to S. Doplicher and J. E. Roberts on the occasion of their 60th birthday}

\begin{abstract}
In conformally invariant quantum field theories one encounters besides the
standard DHR superselection theory based on spacelike (Einstein-causal)
commutation relations and their Haag duality another timelike (''Huygens'')
based superselection structure. Whereas the DHR theory based on spacelike
causality of observables confirmed the Lagrangian internal symmetry picture on
the level of the physical principles of local quantum physics, the attempts to
understand the timelike based superselection charges associated with the
center of the conformal covering group in terms of timelike localized charges
lead to a more dynamical role of charges outside the DR theorem and even
outside the Coleman-Mandula setting. The ensuing plektonic timelike structure
of conformal theories explains the spectrum of the anomalous scale dimensions
in terms of admissable braid group representations, similar to the explanation
of the possible anomalous spin spectrum expected from the extension of the DHR
theory to stringlike d=1+2 plektonic fields.
\end{abstract}\maketitle

\thispagestyle{empty}
\section{Introduction}

Among the oldest and most fruitful concepts in quantum mechanics and quantum
field theory are the spin-statistics connection, the PCT-theorem and the
factorization of the total symmetry into inner- and spacetime-symmetries
\cite{1}. \linebreak Spin\&statistics and PCT were first seen in the formal Lagrangian
quantization approach, whereas the internal symmetry entered particle physics
initially via the phenomenologically motivated approximate isospin symmetry of
nuclear physics and was easily incorporated into the Lagrangian framework in
the form of field/particle multiplets. The DHR-theory \cite{2}, which started
from the properly mathematically formulated causality and spectral principles
for observables of (what become later known as) algebraic QFT \cite{3} and
aimed at the reconstruction of charge carrying (non-observable, superselected)
field operators, finally culminated in the theorem of Doplicher and Roberts
\cite{4}. In this way it became clear that the local quantum physics generated
by a physically admissible field multiplet with a specific internal symmetry
group was already uniquely (after fixing some conventions) characterized by
the observable structure. This de-mystified to a large degree the concept of
internal symmetries by showing a new way to derive the representation category
of compact groups (all compact groups arise in this way) from localization
principles of quantum observables, a quite unexpected connection which has not
yet been fully appreciated by the particle physics community.

Moreover the Spin\&Statistics and TCP issue became inexorable linked with that
of internal symmetry and the original Einstein-causal observable algebra was
reattained as the fixed-point algebras under the compact global ''gauge
group''. Although this picture about internal symmetries confirmed the formal
observations in the Lagrangian quantization setting i.e. there were no
completely unexpected new physical concepts (the innovative power especially
of the DR theory remained on the mathematical side), the superselection
analysis of observable algebras was able to relate hitherto seemingly
unrelated structures and thus lead to a fresh and novel point of view with
different perspectives besides contributing a new mathematical duality theory
on group representations.

The only exceptions were low dimensional field theories ($D<1+3$) where models
were found by special non-Lagrangian methods and where the algebraic methods
led to the more general braid group- instead of permutation group- statistics
\cite{3} for which there are no natural Lagrangian realizations.

Since one of the main localization prerequisites of this theory is the
possibility of compact spacetime localization and since this requirement in
conformal quantum field theories is formally automatically met as a result of
the conformal equivalence of noncompact regions (e.g. wedges) with compact
ones (e.g. double cones or ''diamonds''), the Doplicher Roberts theorem is in
particular applicable to all conformal higher dimensional ($D\geq1+3$) theories.

However it was realized rather early that conformal theories have additional
superselection rules which have a somewhat different conceptual basis and are
intimately related to anomalous scale dimensions. They result from the
structure of the center $Z$ of the conformal covering whose action describes a
timelike rotational sweep and hence they are not accounted for by the DR
theory. In this paper we look for arguments that the coherent subspaces
associated with the conformal covering group are also of local origin i.e.
associated with the representation theory of an algebra with timelike locality
\cite{9}. In fact it was noticed that the ensuing conformal decomposition
theory is nonlocal at spacelike differences \cite{5}, but its timelike
structure remained unexplored.

Only in the very special and atypical D=1+1 conformal theories which permits a
topology preserving interchange between the space- and time-like regions and
which leads to a tensor decomposition into two ''chiral'' lightray theories, a
sufficiently rich family of nontrivial (''minimal'') models (abelian braid
group illustrations with exponential Bose fields were already discussed in
\cite{5}) was later found by Belavin Polyakov and Zamolodchikov \cite{6}; in
fact the chiral version of the conformal central decomposition theory is part
of their ``block-decomposition. The BPZ methods were based on special
algebraic structures which had no counterpart in higher spacetime dimensions.
By emphasizing the charge transport around the compactified Minkowski world
(charge-monodromy) \cite{7} and the related braid group statistics as
expressed in terms of exchange algebras, it was possible to incorporate chiral
quantum field theory into the algebraic setting of superselection sectors i.
e. to place it under a common roof with higher dimensional QFT.

The suggestion to look for timelike braid group commutations in higher
dimensional conformal theories is consistent with the analytic structure of
the two-point function which for scalar fields is%

\begin{align}
\left\langle A(x)A(y)^{\ast}\right\rangle  &  \simeq lim_{\varepsilon
\rightarrow0}\frac{1}{\left[  -\left(  x-y\right)  _{\varepsilon}^{2}\right]
^{\delta_{A}}}\,\\
\left(  x-y\right)  _{\varepsilon}^{2}  &  =\left(  x-y\right)  ^{2}%
+i\varepsilon(x_{0}-y_{0})\nonumber\\
\left\langle A(x)A(y)^{\ast}\right\rangle  &  =e^{2i\delta_{A}}\left\langle
A(y)^{\ast}A(x)\right\rangle \,,\ x>y
\end{align}
where the $\varepsilon$ boundary prescription is just the spacetime version of
the energy-momentum positivity and $\lessgtr$ denote $\mp$timelike
separationes. One observes that for timelike distances the commutation
relation can be at best plektonic\footnote{As it has become costumary in AQFT
``plektonic'' is used for the general (abelian and nonabelian) physically
admissible braid group representation whereas anyonic refers to the abelian
case (which is closer to the standard formulation of QFT).} and certainly not
bosonic/fermionic. But the two-point function does not reveal anything
substantial concerning localization of fields and in particular 2- and 3-point
functions cannot distingush anyonic (abelian) from general plektonic
(nonabelian) timelike braid group structure. The consistency with higher point
functions will be presented in section 3.

It is well known from chiral theories (where distances are lightlike) that the
plektonic superselection structure is inexorably linked to the appearance of
nontrivial central projectors which are the spectral projectors in the
spectral resolution of the abelian generator $Z$ of $center(\widetilde
{SO(4,2)})=\left\{  Z^{n};n\in\mathbb{Z}\right\}  .$ In chiral theories, which
are based on the factorization $\widetilde{S(2,2)}\simeq\widetilde
{SL(2,R)\times SL(2,R)},$ a very good understanding about a one-to-one
relation between algebraic nets of AQFT and conformal equivalence classes of
generating ``field coordinates'' used in standard QFT has been achieved, and
even the problem how to construct pointlike fields from nets of algebras has
received successful attention \cite{10}. There can be no reasonable doubts
that these considerations can be generalized the higher dimensional conformal
case, and in the present paper we will present some consistency arguments to
this effect.

Even though conformal theories are somewhat outside of particle physics proper
(since interactions, although consistent with all other properties of QFT, are
inconsistent with a bona fide zero mass particle structure \cite{8}), they
still are expected to furnish useful illustrations of interacting local
quantum physics.

In the next section we prepare the geometric prerequisites. This material is
well known but we need to remind the reader and to set our notation. In the
same section we also review the expansions with respect to the central
projectors of $Z.$ The core of this paper is section 4 where the consistency
of timelike plektonic structures is discussed within the Wightman framework
and were one can also find some remarks about some concepts which hopefully
will turn out to be important in an algebraic setting.

\section{Conformal Central Decomposition}

According to Wigner the projective aspect of states in quantum theories
requires the action of the universal covering of symmetry groups to act on
state vectors in Hilbert space. Whereas for the Poincar\'{e} group the double
covering explains the phenomenon of halfinteger spin and its relation to Fermi
statistics, the larger conformal group has a much richer infinite covering
symmetry i.e. the center of the conformal covering $\widetilde{SO(4,2)}$ is
generated by one abelian element $Z$ of infinite order\footnote{Actually the
physical conformal group is $SO(d,2)/Z_{2}$, but for our purpose its double
covering is more suitable.}. Corresponding to spacelike $2\pi$ rotation as
compared to the timelike sweep through $\bar{M},$ the related physical
phenomena are somewhat different. Whereas the spatial spin-statistics
connection and associated univalence superselection rule appeared quite early
in the famous work of Wick Wightman and Wigner at the beginning of the 50s
\cite{3} and marked the beginning of the discussion about limitations of the
quantum mechanical superposition principle due to superselection sectors, the
conformal superselection rule which required the setting of local QFT and led
to the temporal conformal decomposition theory, was discovered only twenty
years later \cite{5}. There is a formal similarity between both since whereas
in the case of spin a $2\pi$ rotation in space results in a $e^{2\pi is}$
phase factor on vectors of spin s is related to the spacelike commutation
structure for localized operators, the conformal case permits in addition a
timelike rotational sweep which is associated via the eigenvalues $e^{2\pi
i\delta}$ of $Z$ to the spectrum of anomalous dimensions. Whereas the
connections between the anomalous dimensions with the central phases in full
timelike sweeps and the associated timelike decomposition theory into
superselected sectors of local operators was obtained already in the 70s
\cite{5}, the possible connection (there are as yet no controllable models)
with a timelike braid group structure of charge-carrying fields associated
with timelike commuting observables fulfilling Huygens principle is of a
fairly recent vintage \cite{9} and constitute the main subject of this report.

One reason for this delayed attention to such a fundamental problem is of
course that the required methods have neither a natural place in the
Lagrangian approach, nor are they in reach of the BPZ \cite{6} representation
theoretical methods (e.g. no immediate analog of locally acting
diffeomorphisms beyond the finite parametric conformal group exists) whose
algebraic structure is restricted to chiral theories. As will be seen in the
sequel they are even somewhat outside the formalism of DHR since the issue of
global causality in the presence of a covering of spacetime tends to be more
``dynamical'' than the basically kinematical DHR superselection analysis. The
step to re-derive or incorporate the chiral results into the general setting
of locally generated superselected charges by liberating them from the rather
special diffeomorphism- and loop-group algebras algebras has been achieved in
a series of papers, for the most recent (with references to prior ones) see
\cite{11}.

Although the similarities with D=1+1 in the covering and causality aspects are
helpful, one must also appreciate the differences. The most important
difference is already visible on the classical level when one studies the
characteristic value problem. It is well known that for D%
$>$%
1+1 that data on that part of the light front which constitutes the upper
causal horizon of a wedge region already fully determines the data in the
wedge region, whereas for D=1+1 one needs the data on both lower and upper
horizon to determine the data inside the wedge. The latter fact is of course
intimately related to the D=1+1 decomposition into chiral right and left
movers which for the quantum observables prevails even the presence of
interactions. \ These differences have their counterpart in local quantum
physics \cite{9}.

The content of the present paper aims at showing consistency between the
Boson/Fermion statistics structure of the spacelike based DHR theory and the
appearance of new central decomposition superselection sectors which require
an autonomous role for the timelike region. Whereas the timelike region is the
arena of interactions which in massive interactions has remained impenetrable
to direct investigation, conformal symmetry opens this region to a full
analysis for interacting charge-carrying fields. This phenomenon has no chiral counterpart.

Before we present the timelike braid group structure and the resulting
classification theory for anomalous scale dimensions in the next section, we
will review briefly the known facts about the conformal covering structure and
the decomposition theory of local fields in the remainder of this section.

It is customary to compactify D-dimensional Minkowski space $M$ within a
$(D+2)$-dimensional linear formalism \cite{Lü} with signature $(D,2)$
corresponding to the $SO(D,2)$ group with signature (+----+) where + means
timelike. The surface of the forward light cone is a D+1 dimensional
submanifold $LC^{(d+1)}=\left\{  \xi,\;\xi_{\mu}\xi^{\mu}=0\right\}  $ and the
$D$-dimensional manifold of directions on this surface is the model for the
compactified Minkowski space $\bar{M}.$ The following parametrization which is
also useful for the infinite sheeted covering $\widetilde{M}$ of $\bar{M}$ is
well known ($\tau="$conformal time'')
\begin{align}
\bar{M}  &  =(sin\tau,\mathbf{e},cos\tau),\,\,-\pi<\tau<\pi,\,\mathbf{e}%
^{2}=1\\
\bar{M}  &  \simeq S^{3}\times S^{1}\nonumber
\end{align}
In terms of the $D$-dimensional standard coordinates it reads
\begin{equation}
x^{0}=\frac{sin\tau}{cos\tau+e^{d}},\,\ \vec{x}=\frac{\mathbf{\vec{e}}%
}{cos\tau+e^{d}}%
\end{equation}
where the boundary of $\bar{M}$ correspond to infinite remote points in the
Cartesian coordinates (the usual covering model which one associates with the
standard coordinates is $\bar{M}/Z_{2}$ together with the corresponding group
$\widetilde{SO(D,2)}/Z_{2}$). Since the covering space has the topology
\begin{equation}
\widetilde{M}=(\mathbf{e},\tau)\simeq S^{d-1}\times\mathbb{R},
\end{equation}
the causal dependence region in the global sense of the covering space is the
noncompact complement of the compact spacelike region. In terms of differences
between events ($\mathbf{e},\tau$) and ($\mathbf{e}^{\prime}$,$\tau^{\prime}$)
in $\widetilde{M}$ we have for globally causal relations
\begin{align}
\left|  \tau-\tau^{\prime}\right|   &  >\left|  Arcos(\mathbf{e\cdot
e}^{\prime})\right|  \,\,\pm timelike\label{causal}\\
\left|  \tau-\tau^{\prime}\right|   &  <\left|  Arcos(\mathbf{e\cdot
e}^{\prime})\right|  \,\,\,spacelike
\end{align}
and in a graphical representation\footnote{$M$ looks then like a Penrose
world, except that Penrose does not make the $\bar{M}$ identifications because
his matter content is not invariant in the sense of Huygens priniple.} in
terms of the surface of a D+1 dimensional cylinder one has a tiling in terms
of infinitely many repeated diamond-shaped Minkowski spacetimes with a d-1
dimensional ``infinity'' $\bar{M}\backslash M$ which is spanned by the
backward light cone with apex at $m_{+\infty}=(0,0,0,0,1,\tau=\pi)$
intersected with the forward cone with apex $m_{-\infty}(\tau=-\pi)$ \cite{9}.
As on $S^{1},$ there is no genuine causality concept on $\bar{M}$; algebras
commute whenever the light rays emanated from the localization region of one
do not intersect the other. A glocal notion of causality is however restored
on $\widetilde{M}$ \cite{Se}\cite{Lü}$.$

There exists an economical way to organize the conformal transformations
relative to the Poincar\'{e} subgroup which consists in defining the analogue
of the chiral rotation with the help of the conformal inversion acting on
translations
\begin{align}
R_{\mu}  &  =P_{\mu}+IP_{\mu}I\\
I  &  :x\rightarrow\frac{-x}{x^{2}}\nonumber
\end{align}
The inversion itself is not part of the connected conformal group (except in
free theories), but the product $IP_{\mu}I$ is the generator of the
fractionally acting abelian subgroup corresponding to $x\rightarrow
\frac{x-bx^{2}}{1-2bx+b^{2}x^{2}}$ whereas $R_{\mu}$ generates a kind of
''translation'' analogue which acts as a timelike rotation through the compact
$\bar{M}.$ In fact if one looks at
\begin{align}
&  U_{e}(\tau)=e^{i\tau e\cdot R}\label{circle}\\
&  eR=e^{\mu}R_{\mu},\,\,e^{2}=1,\,\,e_{0}>0\nonumber
\end{align}
one realizes that $U_{e}(\tau)$ in the rest frame is precisely the so called
conformal-time transformation which plays the crucial role in the
compactification and which for $\tau=2\pi$ defines the generator of the center
of $\widetilde{S(D,2)}.$ The advantage of the above formalism is that it
presents the full conformal group by starting from the Poincar\'{e} group
extended by scale transformations and associating only one additional one
parametric subgroup namely the conformal ``time'' rotations; the rest follows
from Lorentz transformations. This makes the topological similarity of
$S^{3}\times S^{1}$ with the well known chiral case analytically very
explicit. In particular the well-known statement that observable chiral fields
on $S^{1}$ have meromorphic analytically continued correlation functions
paases to higher dimensional conformal observables on $\bar{M}\simeq
S^{3}\times S^{1}.$ In this analytic language the cuts of correlations of
charge-carrying fields on the complex extension of $\bar{M}$ disappear in the
transition to the complexification on $\widetilde{M}.$

The basic observation which led the present author et. al. \cite{5} to the
decomposition theory for covariant local Boson/Fermi charge-carrying fields
$F$ was that one obtains quasiperiodic fields on $M$ which remain irreducible
even under global conformal transformations including those involving the
action of the center of the group which has one abelian generator $Z\,$
$center(\widetilde{S(D,2)})=\left\{  Z^{n},n\in\mathbb{Z}\right\}  $
\begin{align}
F(x)  &  =\sum_{\alpha,\beta}F_{\alpha,\beta}(x),\,\,F_{\alpha,\beta}(x)\equiv
P_{\alpha}F(x)P_{\beta}\label{dec}\\
Z  &  =\sum_{\alpha}e^{2\pi i\theta_{\alpha}}P_{\alpha}\nonumber
\end{align}
in terms of central projectors. In a way the existence of this decomposition
facilitates the use of the standard parametrization of Minkowski space
augmented by the quasiperiodic central transformation
\begin{equation}
ZF_{\alpha,\beta}(x)Z^{\ast}=e^{2\pi i(\theta_{\alpha}-\theta_{\beta}%
)}F_{\alpha,\beta}(x)
\end{equation}
and hence one may to a large part avoid the use of the complicated covering
parametrization and its $\widetilde{SO(D,2)}$ transformations which the
unprojected fields $F$ would require. For the latter fields on $\widetilde{M}$
the notation would be insufficient; one also has to give an equivalence class
of path (the number $n\gtrless$0\ of the heaven/hell one is in) with respect
to our copy of $M$ embedded in $\widetilde{M}.$ The projected fields on the
other hand are analogous to sections in a trivialized vectorbundle.

Whereas spacelike causality on $M$ and $\bar{M}$ is not conformally invariant
(only lightlike separations are invariant), the global distinction in
$\widetilde{M}$ between \linebreak positive/negative timelike and spacelike is invariant
and corresponds to the two sign in (\ref{causal}). The attachment of an index
$n$ to the projection in Minkowski spacetime prevents that a pair of points in
$\widetilde{M}$ which was spacelike can become timelike under a transformation
since e.g. it prevents the passing through lightlike infinity via special
conformal transformations $x\rightarrow\frac{x-bx^{2}}{1-2bx+b^{2}x^{2}}$. In
this way we solved the causality paradox \cite{Hort} since it only came about
by forgetting the path dependence which linked the Minkowski ``heavens and
hells'' to our Minkowski space \cite{5}. Instead of the $(\mathbf{e},\tau)$
parametrization (\ref{causal}), we used a function of a pair of conformal
transformations $\sigma(C_{1},C_{2})$ which can be obtained from the quadratic
expression $\sigma(b,x)=1-2bx+b^{2}x^{2}$ \cite{5}\cite{3}.

It was shown \cite{5} via the conformal properties of 3-point functions that
spec\linebreak $\left\{  Z^{n},n\in\mathbb{Z}\right\}  =e^{2\pi i\delta},\delta
\,:an.\,dim.\}.\,$Strictly speaking it is not the dimension but rather the so
called ''twist'' $t=\delta-s$ where $s$ is the spin \cite{5}, but here we
restrict ourselves to bosonic theories.

All above formula in fact remain true in the case D=1+1 if one takes care of
the chiral tensor product structure which leads to a bigger tensor product
center $\widetilde{SO(2,2)}=\left\{  Z_{+}^{n_{+}}\times Z_{-}^{n_{-}%
}\right\}  .$ In that case the D=1+1 fields can be projected by factorizing
double-indexed projectors $P_{\alpha_{+}\alpha_{-}}=P_{\alpha_{+}}^{(+)}\times
P_{\alpha_{-}}^{(-)}$ onto charge sectors which refine the central projectors
i.e. a central projector is a sum over charge projectors$.$ Restricting to one
chiral factor, one finds a lightlike plektonic exchange algebra for double
indexed charge-carrying fields or operators (removing the $\pm$ notation)
\begin{align}
F_{\alpha,\beta}(x)G_{\beta,\gamma}(y)  &  =\sum_{\beta^{\prime}}%
R_{\beta,\beta^{\prime}}^{(\alpha,\gamma)}G_{\alpha,\beta^{\prime}}%
(y)F_{\beta^{\prime},\gamma}(x),\,\,x>y\label{time}\\
F_{a,\beta}G_{\beta,\gamma}  &  =\sum_{\beta^{\prime}}R_{\beta,\beta^{\prime}%
}^{(\alpha,\gamma)}G_{\alpha,\beta^{\prime}}F_{\beta^{\prime},\gamma
},\,\,locF>locG \label{ex}%
\end{align}
i.e. a commutation relation with R-matrices which form a representation of the
infinite braid group. The more general algebraic form (\ref{ex}) of the
exchange algebra in terms of operator algebras instead of fields was derived
in \cite{7}.

Since in higher dimensions only the timelike region has an ordering structure
which is maintained by positive respectively negative central transformations
$Z^{\pm1},$ the exchange relations are are geometrically consistent for the
timelike region in any dimension with $\lessgtr$ meaning positive/negative
timelike. In the next section we will test the consistency of this plektonic
structure with the spacelike bosonic/fermionic commutation relation.

\section{Timelike Decomposition Structure and the Braid Group}

For chiral theories the structural investigations by the methods of algebraic
quantum field theory were proceeded by a good understanding of exchange
algebras in the more standard setting \cite{R-S} of Wightman fields and their
correlation functions. It is reasonable to proceed in the same way for higher
dimensional conformal QFT.

The most powerful tool of Wightman's formulation is provided by the analytic
properties of correlation functions. It is well known that the complexified
Lorentz group may be used to extend the tube analyticity associated with the
physical positive energy-momentum spectrum. The famous BHW theorem \cite{1}
insures that this extension remains univalued in the new complex domain and
the Jost theorem characterizes its real points. Finally spacelike locality
links the various permutations of the position field operators within the
correlation function to one permutation (anti)symmetric analytic master
function which is still univalued. The various correlation functions on the
physical boundary with different operator ordering can be obtained by
different temporal $i\varepsilon$ prescriptions.

Complexifying the scale transformations, the conformal correlations can be
extended into a still bigger analyticity region which even incorporates
``timelike Jost points'' but trying to find a master function which links the
various orders together fails in the presence of fields with anomalous
dimensions and remains restricted to fields which live on the compactification
$\bar{M}.$ The latter are the analogs of chiral observables, except that apart
from (composite) free fields one does not have algebraic examples since
Virasoro- and Kac-Moody algebras do not exist in higher dimensions.

The analytic timelike structure of 3-point functions suggest that the
permutation group should be replaced by the more general braid group. The
global timelike ordering structure on the covering $\widetilde{M}$ is the
prerequisite; without this ordering one can only have the more special
permutation group commutations since the exchange and its inverse can then be
continuously deformed into each other.

A plektonic (general braid-type) charge structure which is only visible in the
timelike region would immediately explain the appearance of a nontrivial
timelike center and the spectrum of anomalous dimension. It would kinematize
conformal interactions and reveal conformal QFT as basically free theories if
it would not be for that part of interaction which sustains the timelike
plektonic structure. Of course the situation trivializes if the theory has no
anomalous dimensions and nontrivial components. Analogous to \cite{3} (remarks
at end of section V.4) we conjecture that this characterizes interaction free
conformal theories which are generated by free fields\footnote{Note that this
conjecture would be wrong in D=1+1 since from selfdual lattice construction on
current algebras one obtains models without nontrivial sectors which are
different from free fields.}. What makes this issue somewhat complicated is
the fact that contrary to chiral theories we do not have a single nontrivial
example because this issue is neither approachable from the representation
theory of known infinite dimensional Lie-algebras nor from the formal
euclidean functional integral method. The remaining strategy is to show
structural consistency of the spacelike local- with the conjectured timelike
plektonic- structure and to find a new construction method (non
energy-momentum tensor- or current- algebra based, non-Lagrangian). Here we
are mainly concerned with consistency arguments and in the following we will
comment how local/plektonic on-vacuum relations between two fields can be
commuted through to a generic position.

Assume for simplicity as before that we are in a ``minimalistic'' situation
where the field theory has no internal symmetry group\footnote{The general
exchange algebra relations with group algebra valued R-matrices have been
elaborated by K-H Rehren (private communication).}, but that the fields can be
given ``timelike'' charge indices $\alpha,\beta,\gamma..$ and their conjugates
$\bar{\alpha},\bar{\beta},\bar{\gamma}....$ resulting from projectors on
charge spaces so that the decomposition is as in the chiral case where the
charge projectors with the same phase factors $e^{2\pi i\delta}$ constitute a
refinement of the central projectors. Clearly $\alpha$ and its conjugate
$\bar{\alpha}$ contribute to the same central projector. In fact we may take
over a substantial part of the formalism and concepts of \cite{R-S} if one
replaces the chiral translation+dilation augmented with the circular rotation
generator $L_{0}$ by the spacetime symmetry group which leaves the timelike
infinite point fixed (Poincar\'{e}+dilations) extended by the generator of
conformal time $R_{0}$ instead of the chiral $L_{0}.$ One would of course also
have to change the title of the old paper from ``Einstein causality and Artin
braids'' to ``Huygens causality and Artin braids'' referring to the timelike
ordering for which the conformal observables fulfill the Huygens principle of
vanishing commutators. The ``on-vacuum'' structure of commutation relations
follows from the structure of the conformal 3-point functions (here the
$F,G,H$ fields are not observable fields but are as the F,G of the previous
section)
\begin{align}
&  \left\langle H^{\ast}(x_{3})G(x_{2})F(x_{1})\right\rangle =c_{FGH}\frac
{1}{\left[  -(x_{12})_{\varepsilon}^{2}\right]  ^{\delta_{3}}}\frac{1}{\left[
-(x_{13})_{\varepsilon}^{2}\right]  ^{\delta_{2}}}\frac{1}{\left[
-(x_{23})_{\varepsilon}^{2}\right]  ^{\delta_{1}}}\\
&  \delta_{1}=\frac{1}{2}(\delta_{F}+\delta_{H}-\delta_{G}),\,\delta_{2}%
=\frac{1}{2}(\delta_{G}+\delta_{H}-\delta_{F}),\,\delta_{3}=\frac{1}{2}%
(\delta_{F}+\delta_{G}-\delta_{H})\nonumber
\end{align}
where the $\varepsilon$-prescription was explained in the introduction. For
spacelike and timelike distances one concludes
\begin{equation}
G(x_{2})F(x_{1})\Omega=\left\{
\begin{array}
[c]{c}%
F(x_{1})G(x_{2})\Omega,\,\left(  x_{2}-x_{1}\right)  ^{2}<0\\
e^{\pi i(\delta_{G}+\delta_{F})}Z^{\ast}F(x_{1})G(x_{2})\Omega,\,\left(
x_{2}-x_{1}\right)  ^{2}>0,\,\left(  x_{2}-x_{1}\right)  _{0}>0
\end{array}
\right.
\end{equation}
since this relation is valid on all quasiprimary composites $H.$ They consist
of the equal point limit of the associated primary $H_{min}$ (lowest scale
dimension operator in the same charge class) multiplied with a polynomial in
the observable field. These composites applied to the vacuum form a dense set
in the respective charge sector\footnote{With a bit more work and lengthier
formulas one can avoid the colliding point limit and use correlation functions
containing 3 charged fields and an arbitrary number of neutral observable
fields. The dependence on the observable coordinates is described by a
rational function on $\bar{M}.$} and hence the on-vacuum formula is a
consequence of the structure of 3-point functions. The spacelike local
commutativity off-vacuum is consistent with that on-vacuum since for $y$
timelike with respect to the spacelike pair $x_{1},x_{2}$ we have ($c_{F}%
$=superselected charge of $F$)
\begin{align}
P_{\alpha}F(x_{1})G(x_{2})H(y)\Omega &  =\sum_{\beta}P_{\alpha}F(x_{1}%
)P_{\beta}G(x_{2})H(y)\Omega\nonumber\\
&  =\sum_{\beta}P_{\alpha}F(x_{1})P_{\beta}e^{i\pi(\delta_{G}+\delta
_{H}-\delta_{\beta})}H(y)G(x_{2})\Omega\\
&  =\sum_{\beta\beta^{\prime}}R_{\beta\beta^{\prime}}^{(\alpha\gamma)}%
(c_{F},c_{G})e^{i\pi(\delta_{G}+\delta_{H}-\delta_{\beta})}P_{\alpha
}H(y)P_{\beta^{\prime}}F(x_{1})P_{\gamma}G(x_{2})\Omega\nonumber
\end{align}
and therefore the off-vacuum vanishing of the $F$-$G$ commutator is consistent
with the on-vacuum vanishing of this commutator if there holds a certain
relation between $R(c_{F},c_{G})$ and $R(c_{G};c_{F})$ which is identically
fulfilled for $c_{F}=c_{G}$. Similarly one does not run into inconsistencies
if one tries to obtain a timelike off-vacuum $F$-$G$ situation from the
on-vacuum placement by commuting through a $H$ which is spacelike to the
timelike $F$-$G$ pair
\begin{align}
&  P_{\alpha}F(x_{1})G(x_{2})H(y)\Omega=P_{\alpha}H(y)F(x_{1})G(x_{2}%
)\Omega=P_{\alpha}H(y)e^{i\pi(\delta_{G}+\delta_{H}-\delta_{\beta})}%
G(x_{2})F(x_{1})\Omega\nonumber\\
&  \,\,\,=\sum_{\beta}R_{\beta\beta^{\prime}}^{(\alpha\gamma)}(c_{F}%
,c_{G})P_{\alpha}G(x_{2})P_{\beta^{\prime}}F(x_{1})P_{\gamma}H(y)\Omega
\nonumber \\
& \,\,\,=\sum_{\beta\beta^{\prime}}R_{\beta\beta^{\prime}}^{(\alpha\gamma)}%
(c_{F},c_{G})P_{\alpha}G(x_{2})P_{\beta^{\prime}}H(y)F(x_{1})\Omega
\end{align}
where in the second line we commuted $F$ through $G$ before trying to bring
both to the vacuum. Since their is no rule to commute the $P_{\alpha}%
G(x_{2})P_{\beta^{\prime}}$ with $P_{\beta^{\prime}}H$ for $\left(
x_{2}-y\right)  ^{2}<0,$ there is no way to get to the same $HGF$ order as in
the first line and hence no consistency relation to be checked. The absence of
rules for spacelike commutations for projected fields protects the formalism
to run into inconsistencies.

Let us also briefly look at the compatibility of the timelike plektonic
structure with the conformal structure of the 4-point function of 4 identical
hermitean fields%

\begin{align}
&  W(x_{4},x_{3},x_{2},x_{1}):=\sum_{\gamma}\left\langle F(x_{4}%
)F(x_{3})P_{\gamma}F(x_{2})F(x_{1})\right\rangle \\
&  =\left[  \frac{x_{42}^{2}x_{31}^{2}}{(x_{43})_{\varepsilon}^{2}%
(x_{32})_{\varepsilon}^{2}(x_{21})_{\varepsilon}^{2}(x_{14})_{\varepsilon}%
^{2}}\right]  ^{\delta_{A}}\sum_{\gamma}w_{\gamma}(u,v),\,\,\nonumber\\
&  \,\,\,u=\frac{x_{43}^{2}x_{21}^{2}}{(x_{42})_{\varepsilon}^{2}%
(x_{31})_{\varepsilon}^{2}},\,\,v=\frac{x_{32}^{2}x_{41}^{2}}{(x_{42}%
)_{\varepsilon}^{2}(x_{31})_{\varepsilon}^{2}}\nonumber
\end{align}
Whereas the spacelike commutations leads to functional relations for
\linebreak $w=\sum_{\gamma}w_{\gamma}(u,v)$ with the exchange of two fields causing a
rational transformation of the $u,v$ (apart from multiplying the $w$ by
rational $u,v$ factors)$,$ the timelike commutation of the off-vacuum fields
produces rational transformation together with monodromy R-matrix mixing of
the $\gamma$-components \cite{9} (in addition to multiplying the $w$ with
noninteger powers of u and v which depend on the scale dimension $\delta_{F}%
$). Despite some similarities with the chiral case, the dependence of
$w_{\gamma}$ on two cross ratios probably requires the use of more elaborate
techniques than the hypergeometric formalism which is sufficient for the
chiral one variable cross ratio dependence Here we will not pursue this matter.

In order to incorporate these observations on correlation functions into the
algebaic approach one should start from a theorem \cite{Lo} which shows that a
locally conformal field net on $M$ allows a natural extension $\widetilde
{\mathcal{F}}$ to a Haag dual net on $\widetilde{M}.$ The difficult step is to
prove that there exists a nontrivial (observable) subalgebra $\mathcal{A}$ on
$\bar{M}.$ The geometric complement of a double cone $\mathcal{O}$ which is
relevant for Haag duality of $\mathcal{A}$ consists of all points on $\bar{M}$
which are not lightlike to $\mathcal{O}$ \cite{3}.

An attempt to show the existence of $\mathcal{A}$ by modular method shows the
difficulty. Consider the inclusion
\begin{equation}
(\mathcal{F}(\mathcal{V}_{+}^{t})\subseteq\mathcal{F(V}_{+}\mathcal{)},\Omega)
\end{equation}
where $\mathcal{F(V}_{+}^{t})$ is the forward lightcone algebra shifted upward
in time by $t$. One easily checks that this inclusion is modular, i.e. that
the modular group of $\mathcal{F(V}_{+}\mathcal{)}$ (the dilation group) in
one direction compresses $\mathcal{F}(\mathcal{V}_{+}^{t})$ further into
$\mathcal{F(V}_{+}\mathcal{)}.$ As a result the relative commutant
\begin{equation}
\mathcal{F}(\mathcal{V}_{+}^{t})^{\prime}\cap\mathcal{F(V}_{+}\mathcal{)}%
\end{equation}
together with the time translation and dilation turns out to define a bosonic
net on the timelike line. The application to the vacuum generates the vacuum
sector $H_{0}$ (by definition) and the covariantized net (using Poincar\'{e}
transformations) of relative commutants if nontrivial, could serve as
definition of the conformal observable net $\mathcal{A}$ on $\bar{M}$. There
is also the net $P_{0}\mathcal{\tilde{F}}P_{0}\,$\ which contains
$\mathcal{A}$ and has the same modular group. The consistency of the above
timelike braid group structure would suggest that these two nets are equal. A
triviality of $\mathcal{A}$ actually appears quite pathological, but
ultimately this problem of existence of nontrivial anomalous dimension has to
be solved by constructive examples.

By cutting $\bar{M}$ open at $\bar{M}\backslash M$ one looses Haag duality,
but one regains it together with a new net after redefinng double cone
algebras as intersections of the forward with the (time-shifted) backward
light cone.which amounts to a timelike dualization. The mechanism, which
involves diluting the net at infinity and offsetting this by making finitely
localized doublr cones bigger, has been nicely explained for free fields in
\cite{His}. This sort of situation where points or subsets are cut out from a
localization region of a net is much better understood in chiral theories.
There the above situation corresponds to punching a hole into the circle (say
at $\infty$) in which case the new net lost the conformal rotation and only
retains translation and dilation. One then recovers Haag duality as well as
full conformal invariance (with an $L_{0}$ with a different low-lying
spectrum) by suitable extending those algebras in the net which do not contain
$\infty.$ The first observation was made in \cite{Bu} and extended by a more
algebraic analysis, including a partial classification of all such extensions
in \cite{Lo}. One expects that a similar construction in the higher
dimensional case will confirm the compatibility between the spacelike DHR
structure and the present ideas on the level of AQFT.

\section{Outlook}

If one wants to use constructions of chiral models as a guide for higher
dimensional conformal models, one must avoid ideas which are obviously limited
to low dimensions, as representation theory of the diffeomorphism- (Virasoro
algebra) or loop-groups (current algebras). Rather one should use the
spacelike (permutation group statistics) and timelike (braid group) structure
in the process of classifying and constructing models.

For observable fields the correlation functions are meromorphic on $M$ and
rational on $\bar{M}$ as functions of the Poincar\'{e} invariants. But past
experience shows that to base a construction on the (linear) properties of
Wightman functions does not really work because the nonlinear positivity
requirements from quantum theory are not controllable in such an approach.

All successful low dimensional model constructions start with concrete
operators in Hilbert space and keep the positivity under control throughout
the whole construction procedure. But even having opted for operator methods,
one still faces the question of whether one should first aim for the
observable algebras and follow the dichotomy of observables-charged fields or
aim directly at the latter. The division into observables/field algebras is
useful for structural investigations and for situations where mathematicians
already have studied algebras (loop groups, Virasoro-diffeomorpisms,..) which
are candidates for observable algebras.

Which objects are more fundamental, observable- or (superselected)
\linebreak field-algebras? This kind of question is a bit reminiscent of what was first
the hen or the egg. From a historical point the fields were there
first\footnote{In fact observable algebras for free fields obtained as the
fixed point subalgebra under the action of internal symmetry groups tend to
gave a more complicated analytical structure than that of the algebra
generated by the free field itself.} before Haag realized already at the end
of the 50$^{ies}$ almost single-handed that it would be a good idea to view
fields in their role of charge-carrying operators as representation
theoretical objects carrying generalized superselected charge. This thought
was extremely fruitful and led 10 years later to the DHR approach and another
20 years later to the DR-theorem. It provided structural insight into the
inner workings of local quantum physics which Lagrangian QFT was unable to
unravel and although it was not designed to lead to instant predictions, it
became a valuable long term investment into QFT. Logically the central
position in the structural analysis belongs to the observable algebras.

I would like to advocate the thesis that for higher dimensional conformal
theories the best constructive strategy is to take the most advanced
mathematical and conceptual tools and return to the old program of
constructing charged fields directly. It appears to me more natural to explain
the rather complicated quantization phenomena of observables (e.g. the
Friedan-Qiu-Shenker c-quantization) in terms of the conceptually simpler
quantization which is inherent in the Makov traces on the braid group. I am
convinced that such an approach exists and that the Tomita-Takesaki modular
theory will play an important role. Relations to the isomorphism between anti
de Sitter and conformal spacetime as well as to perturbative attempts
(conformal supersymmetric Yang-Mills models) can be found in \cite{9}.

The power of the modular localization method is evidenced in recent approaches
involving ``polarization-free-generators'' to low dimensional particle physics
problems \cite{Polar}\cite{Mu}.

Another curious aspect of the present ideas is the very radical way theories
with braid group structure violate the Coleman-Mandula (C-M) theorem
\cite{19}. Braid group structures cannot be encoded into a multiplicities with
a group like action which then factorizes with the spacetime actions of the
conformal symmetry (as in chiral current algebra representations). The
violation in low-dimensional models (chiral models, massive factorizing D=1+1
models) which do not fit the prerequisites of the C-M theorem was of course
well known to those authors, but it seems that everybody expected that this
could at best occur in D=1+2 massive plektonic models but is excluded in
$D>1+2$ theories. The present work suggests that higher dimensional conformal
theories with anomalous dimensions not only do \textit{not} satisfy the
particle prerequisites \cite{8}, but also violate the spacetime/internal
factorization of symmetries predicted by the C-M theorem (even after it was
adapted to supersymmetries). The C-M prohibition of nontrivial amalgamations
of internal and spacetime symmetries applies very much to the DR internal
symmetries, but the charge fusion symmetries behind the central projectors in
conformal theories in any spacetime dimension is definitely outside the C-M
realm. It is not completely excluded that this has consequences for non group
like regularities even in massive theories, since there are no exact
nonabelian flavor symmetries in nature. But presently one has no idea of how
and by what means conformal theories could be related to theories describing
scattering of massive particles.

All this shows that the DHR superselection sector structure has come a long
way, and the statement that braid group structures are excluded in higher
dimensions without any further qualification seem to be on its way out. It
appeares that there is a new dynamical role for an extension of the idea of
superselected charges of which conformal theories are a foreboding.

I am indebted to K.-H. Rehren for valuable suggestions and I also would like
to acknowledge that I learned from Detlev Buchholz that some years ago he also
had ideas about possible braid group structures at timelike distances.

\end{document}